\documentclass{article}
\usepackage{PRIMEarxiv}

\usepackage{mwe} % to get dummy images
\usepackage{comment} % added by me
\usepackage{amsmath} % added by me
\usepackage{amsfonts} % added by me

\usepackage[utf8]{inputenc} % allow utf-8 input
\usepackage[T1]{fontenc}    % use 8-bit T1 fonts
\usepackage{hyperref}       % hyperlinks
\usepackage{url}            % simple URL typesetting
\usepackage{booktabs}       % professional-quality tables
\usepackage{amsfonts}       % blackboard math symbols
\usepackage{nicefrac}       % compact symbols for 1/2, etc.
\usepackage{microtype}      % microtypography
\usepackage{lipsum}
\usepackage{fancyhdr}       % header
\usepackage{graphicx}       % graphics
\graphicspath{{media/}}     % organize your images and other figures under media/ folder

\title{CT-3DFlow : Leveraging 3D Normalizing Flows for Unsupervised Detection of Pathological Pulmonary CT scans}

\author{
  Aissam Djahnine\\
  Claude Bernard University Lyon 1\\
  Philips Health Technology Innovation \\
  Paris, France \\
  \texttt{aissam.djahnine@philips.com} \\
  \And
  Alexandre Popoff \\
  Philips Health Technology Innovation \\
  Paris, France \\
  \texttt{alexandre.popoff@philips.com} \\
  \And
  Emilien Jupin-Delevaux \\
  Department of Radiology\\
  Hospices Civils de Lyon \\
  Lyon, France \\
  \texttt{emilien.jupin-delevaux@chu-lyon.fr} \\
  \And
  Vincent Cottin \\
  Claude Bernard University Lyon 1 \\
  National Reference Center for Rare Pulmonary Diseases, Louis Pradel Hospital \\
  Lyon, France \\
  \texttt{vincent.cottin@chu-lyon.fr} \\
  \And
  Olivier Nempont \\
  Philips Health Technology Innovation \\
  Paris, France \\
  \texttt{olivier.nempont@philips.com} \\
  \And
  Loic Boussel \\
  Claude Bernard University Lyon 1 \\ 
  Department of Radiology\\
  Hospices Civils de Lyon \\
  Lyon, France \\
  \texttt{loic.boussel@chu-lyon.fr} \\
}

\begin{document}
\maketitle

\begin{abstract}
Unsupervised pathology detection can be implemented by training a model on healthy data only and measuring the deviation from the training set upon inference, for example with CNN-based feature extraction and one-class classifiers, or reconstruction-score-based methods such as AEs, GANs and Diffusion models. Normalizing Flows (NF) have the ability to directly learn the probability distribution of training examples through an invertible architecture. We leverage this property in a novel 3D NF-based model named CT-3DFlow, specifically tailored for patient-level pulmonary pathology detection in chest CT data. Our model is trained unsupervised on healthy 3D pulmonary CT patches, and detects deviations from its log-likelihood distribution as anomalies. We aggregate patches-level likelihood values from a patient's CT scan to provide a patient-level 'normal'/'abnormal' prediction. Out-of-distribution detection performance is evaluated using expert annotations on a separate chest CT test dataset, outperforming other state-of-the-art methods.\\
\end{abstract}

\keywords{Unsupervised Anomaly Detection \and Normalizing Flow \and GLOW \and Pulmonary CT}

\section{Introduction}

Normalizing flow (NF) models, such as NICE\cite{dinh_nice_2015}, RealNVP\cite{dinh_density_2017}, and GLOW\cite{kingma_glow_2018} learn the distribution of the observed data by transforming it into a tractable distribution using invertible and differentiable mappings. This allows for exact log-likelihood computation during inference, and anomaly detection methods based on normalizing flows, such as DifferNet \cite{rudolph_same_2021}, CFLOW-AD \cite{gudovskiy_cflowad_2022}, and FastFlow \cite{yu1_fastflow_2021} have achieved high performance on industrial datasets. Although it is possible to apply normalizing flows (NF) on both images and CNN-extracted features, at the risk of ignoring the 3D context offered by volumetric data. Applying NF models on 3D data is possible but raises technical challenges. In theory, a 3D NF model would need to be trained on whole 3D volumes to learn the distribution of normal scans, and the log-likelihood of a sample scan would then be used to classify it as pathological or not. The corresponding memory requirements would however be far superior to 2D approaches, making training more challenging. To our knowledge, advancements in 3D anomaly detection models are scarce. The PET-3DFLOW model \cite{xiong_pet3dflow_2023} operates at the feature level via an encoder-NF-decoder scheme, and computes an anomaly score based on a weighted combination of negative log-likelihood and reconstruction error, producing anomaly maps via reconstruction --- a potentially sub-optimal method \cite{ravanbakhsh_abnormal_2017,schlegl_unsupervised_2017,kingma_auto-encoding_2022, schlegl_unsupervised_2017,schlegl_f-anogan_2019}. 
We propose here a novel approach for anomaly detection and localization in chest CT scans using a 3D patch-based normalizing flow model, whose architecture is a 3D extension of GLOW \cite{kingma_glow_2018}.\\

\section{Methods}

\begin{figure}[htpb]
  \centering
  \includegraphics[width=\linewidth]{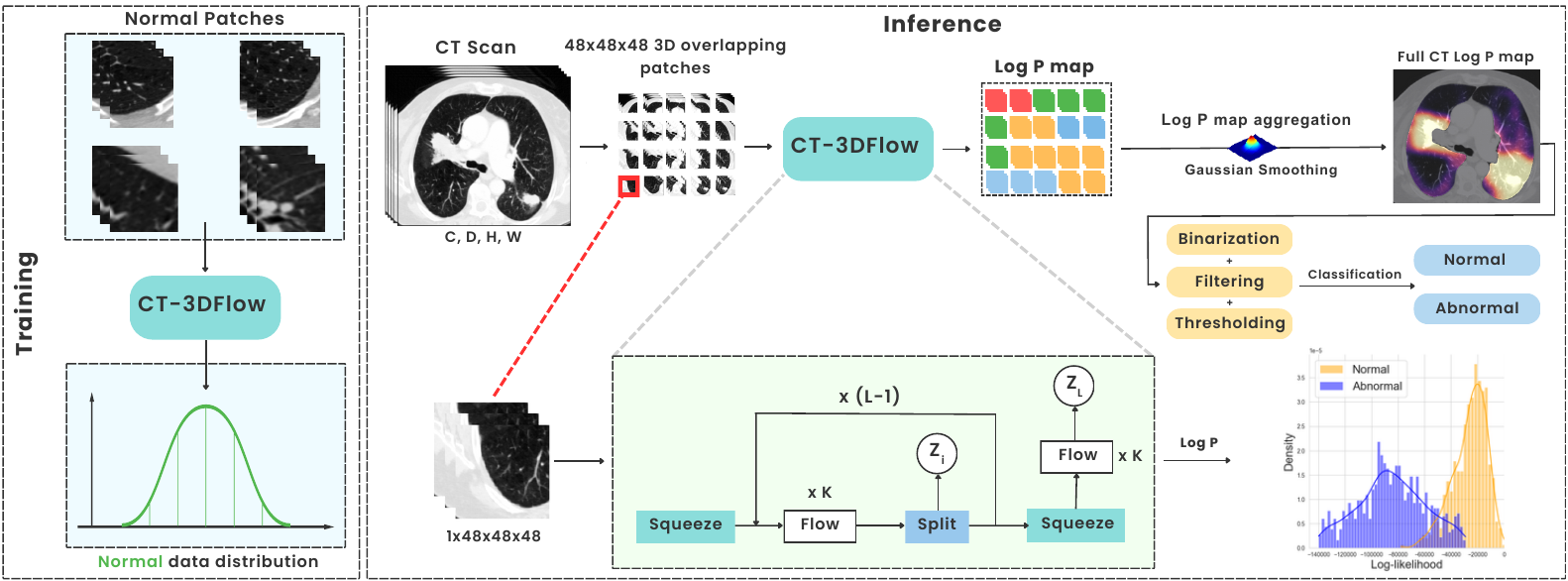}
  \caption{Overview of the CT-3DFlow anomaly detection framework. The model, whose architecture is shown in the middle, is trained on normal 48x48x48 CT sub-volumes. During inference, overlapping patches are processed and aggregated to generate a full CT Log P map, which is binarized, filtered, and thresholded for classification.\\}
  \label{fig:glow_architecture}
  \vspace{-0.5cm}
\end{figure}

We use a dataset of 822 patients for which 3D chest CT scans were obtained from hospital data at a single institution (Hospices Civils de Lyon). It consists of a total of 570 normal and 252 abnormal CT scans (470 scans for training, 111 scans for the validation set, and 291 scans for the test set). These scans are pre-processed by resampling to 2 mm resolution, automatically segmenting the lung in 3D with DL \cite{si-mohamed_automatic_2022} to guide patch-selection, clipping voxel values to the clinically relevant range of $[-1020 HU, +200 HU]$, and normalizing them to $[-0.5, +0.5]$.\\

After training a NF model on 500,000 normal 3D CT 48x48x48-sized patches, we predict abnormalities, consisting of four distinct chest-related abnormalities carefully selected by an expert radiologist based on the prevalence of the most common pulmonary pathologies, at patient-level in a sample CT scan by applying the model on 48x48x48 patches spanning the whole volume (with 10-voxel overlap). Each patch gives a log-likelihood value (indicated as Log P) which serves as a proxy for an anomaly score. These Log P values are aggregated with gaussian smoothing to give a full CT Log P map, which is post-processed to give a binary patient prediction. This involves binarization, filtering, and thresholding. The optimal threshold value $T$ (between $0.5$ and $20$ $cm^{3}$) being obtained from the ROC curve on the validation set (50 normal and 61 abnormal). Finally, we use our test set (50 normal, 191 abnormal) to assess the method's performance. \\

In our experiments, we used a GLOW-based NF architecture which we adapted to 3D with $L=4$ blocks, each one having $K=64$ flows. Each flow consists in an activation normalization layer, an invertible 1x1x1 convolution and an affine coupling layer. The network is trained for 50,000 iterations on 2 NVIDIA A100 SXM4 GPUs, using the maximum likelihood estimation objective with a batch size of 10, using Adam optimizer ($lr=10^{-4}$, weight decay $10^{-5}$). We compare, on a test set, our CT-3DFlow model against several reconstruction-based models (autoencoder AE, and variational autoencoder VAE), GAN-based models, NF-based models (DifferNet, FastFlow, CFLOW-AD), and one diffusion-based model (AnoDDPM). We adopted Area Under the Receiver Operating Characteristic curve (AUROC), F1-score (F1), and accuracy (ACC) as metrics to evaluate the patient-level classification performance of our model.\\

\section{Results and Perspectives}

As shown on Table~\ref{tab:results}, the CT-3DFLow model outperforms the state-of-the-art methods by a noticeable margin in terms of AUC, F1 and ACC. However, further validation of the aggregation methods is required due to the inherent differences between 2D and 3D approaches. Our 3D patch-based NF model demonstrates the superiority of 3D flow-based model over state-of-the-art 2D methods, while mitigating the problems associated to full volume normalizing flows. Further work to generalize this promising approach on other modalities and different organs is envisioned in the future.\\

\begin{table}
 \caption{Comparison of patient-level classification performance with several anomaly detection methods.}
  \centering
  \begin{tabular}{lccc}
    \toprule
    \textbf{Method} & \textbf{AUROC} & \textbf{F1} & \textbf{Accuracy} \\
    \midrule
    AE \cite{kopcan_anomaly_2021}  & 0.608  & 0.538  & 0.556 \\
    VAE \cite{kingma_auto-encoding_2022} & 0.647  & 0.607  & 0.598 \\
    f-AnoGan \cite{schlegl_f-anogan_2019}  & 0.821  & 0.783  & 0.765  \\
    GANomaly \cite{akcay_ganomaly_2019}  & 0.772  & 0.74  & 0.694  \\
    DifferNet \cite{rudolph_same_2021}  & 0.901  & 0.861  & 0.843 \\
    FastFlow \cite{yu1_fastflow_2021}  & 0.942  & 0.912  & 0.903  \\
    CFLOW-AD \cite{gudovskiy_cflowad_2022} & 0.732  & 0.739  & 0.706  \\
    AnoDDPM \cite{wyatt_anoddpm_2022} & 0.924 & 0.891  & 0.873 \\
    \textbf{Ours} & \textbf{0.952} & \textbf{0.94} & \textbf{0.924} \\
    \bottomrule
  \end{tabular}
  \label{tab:results}
\end{table}

%Bibliography
%\bibliographystyle{unsrt}  
%\bibliography{main}  

\end{document}